# Strong Peak in $T_c$ of $Sr_2RuO_4$ Under Uniaxial Pressure

Alexander Steppke[1,2,*,¶], Lishan Zhao[1,2,¶], Mark E. Barber[1,2,¶], Thomas Scaffidi[3], Fabian Jerzembeck[1], Helge Rosner[1], Alexandra S. Gibbs[2,4], Yoshiteru Maeno[5], Steven H. Simon[3,†], Andrew P. Mackenzie[1,2,‡], Clifford W. Hicks[1,§]
15 July 2016

[1]Max Planck Institute for Chemical Physics of Solids, Nöthnitzer Str. 40, 01187 Dresden, Germany
[2]Scottish Universities Physics Alliance (SUPA), School of Physics and Astronomy, University of St. Andrews, St. Andrews KY16 9SS, United Kingdom
[3]Rudolf Peierls Centre for Theoretical Physics, Oxford, 1 Keble Road, OX1 3NP, United Kingdom
[4]ISIS Facility, Rutherford Appleton Laboratory, Chilton, Didcot OX11 OQX, United Kingdom
[5]Department of Physics, Graduate School of Science, Kyoto University, Kyoto 606-8502, Japan

[*]steppke@cpfs.mpg.de  [†]steven.simon@physics.ox.ac.uk  [‡]mackenzie@cpfs.mpg.de
[§]hicks@cpfs.mpg.de

[¶]These authors contributed equally to the experimental work described in this paper.

December 2016

$Sr_2RuO_4$ is an unconventional superconductor that has attracted widespread study because of its high purity and the possibility that its superconducting order parameter has odd parity. We study the dependence of its superconductivity on anisotropic strain. Applying uniaxial pressures of up to ~1 GPa along a ⟨100⟩ direction ($a$-axis) of the crystal lattice results in $T_c$ increasing from 1.5 K in the unstrained material to 3.4 K at compression by ≈0.6%, and then falling steeply. Calculations give evidence that the observed maximum $T_c$ occurs at or near a Lifshitz transition when the Fermi level passes through a Van Hove singularity, and open the possibility that the highly strained, $T_c$=3.4 K $Sr_2RuO_4$ has an even- rather than an odd-parity order parameter.

The formation of superconductivity by the condensation of electron pairs into a coherent state is one of the most spectacular many-body phenomena in physics. Initially, all known superconducting condensates were of the same basic class, in which electrons paired into spin-singlet states, forming condensates of even parity whose phase $\phi$ is independent of wave vector $\mathbf{k}$ [1]. Condensates of this form are insensitive to the presence of non-magnetic scattering, and so are easier to observe in materials grown with standard levels of disorder. In the last three decades, a richer and more exciting picture has emerged. In the growing number of known unconventional superconductors, both the phase and amplitude of the condensate order parameter have strong $\mathbf{k}$ dependence. Unconventional superconductors can have both even and odd parity, and are sensitive to the presence of disorder [2, 3]. These materials give a unique opportunity to study the collective physics of interacting electrons and the mechanisms by which the condensation from the normal metallic state occurs. However, considerable material and experimental challenges must be overcome.

The subject of the research described in this paper, $Sr_2RuO_4$ (transition temperature $T_c \approx 1.5$ K) [4], is the most disorder-sensitive of all known superconductors [5]. However the stringent requirements this places on material purity also bring advantages. The long mean free paths of $\sim 1$ $\mu$m that are required to observe its superconductivity in the clean limit have also enabled extensive studies of its normal state via the de Haas-van Alphen effect [6]. This work, combined with angle-resolved photoemission experiments [7] and electronic structure calculations [8, 9, 10], has led to a detailed understanding of the quasi-2D Fermi surface topography, and the effective masses of the Landau Fermi liquid quasiparticles which pair to form the superconducting condensate.

However, in spite of over two decades of work, the superconducting order parameter is not known with certainty. Soon after the discovery of the superconductivity, the similarity of the

Landau parameters of $Sr_2RuO_4$ to those of the famous $p$-wave superfluid $^3$He led to the proposal that it might be an odd-parity superconductor with spin-triplet $p$-wave pairing [11]. Knight shift measurements [12, 13] and, recently, proximity-induced superconductivity in epitaxial ferromagnetic $SrRuO_3$ layers [14] provide strong evidence for triplet pairing. Muon spin rotation [15] and Kerr rotation [16] experiments point to time reversal symmetry breaking at $T_c$, and tunneling spectroscopy to chiral edge states [17]. Josephson interferometry indicates the presence of domains in the superconducting state and gives evidence for odd parity [18, 19]. In combination, these observations suggest the existence of a chiral, spin-triplet superconducting state with an order parameter of the form $p_x \pm ip_y$. Although the edge currents predicted for chiral $p$-wave order are not seen [20, 21, 22], there are proposals to explain why these might be unobservably small in $Sr_2RuO_4$ [23, 24, 25, 26]. More difficult to explain in the context of spin-triplet pairing is why the upper critical field $H_{c2}$ for in-plane fields is first-order at low temperatures [27] and smaller than predictions for orbital limiting based on anisotropic Ginzburg-Landau theory [28]. More complete reviews of the superconductivity of $Sr_2RuO_4$ and arguments for and against various order parameters can be found in Refs. [29, 30, 31, 32].

The electronic structure of $Sr_2RuO_4$ is relatively simple compared with those of many unconventional superconductors. Its Fermi surfaces are known with accuracy and precision [6] and it shows good Fermi liquid behavior in the normal state [33]. Therefore, gaining a full understanding of the superconductivity of $Sr_2RuO_4$ is an important challenge and a benchmark for the field. An approach not extensively explored so far is to perturb the underlying electronic structure as far as possible from its native state and observe the effects on the superconductivity. Partial substitution of La for Sr [34,35] and epitaxial thin film growth on lattice-mismatched substrates [36] have both been used to push one of the Fermi surface sheets of $Sr_2RuO_4$ through a Lifshitz transition, *i.e.* a topological change in the Fermi surface, and an associated Van Hove singularity (VHS) in the density of states. This is a major qualitative change in the electronic structure, and it would be interesting to see how the superconductivity responds. However, the disorder sensitivity of the superconductivity of $Sr_2RuO_4$ is so strong that it was not possible to do either experiment in a sufficiently clean way for any superconductivity to survive.

In principle, uniaxial pressure has the potential for tuning the electronic structure of $Sr_2RuO_4$ without introducing disorder and destroying the superconductivity. Pressure applied along a ⟨100⟩ lattice direction, lifting the native tetragonal symmetry of $Sr_2RuO_4$, has been shown to increase the bulk $T_c$ to at least 1.9 K [37]. There are hints that $T_c$~3 K in pure $Sr_2RuO_4$ is achievable with lattice distortion [38, 39], however it has only been seen locally, which complicates determination of its origin and properties. By extending the piezoelectric-based compression techniques introduced in Ref. [37] to achieve much higher compressions, we demonstrate in this work the existence of a well-defined peak in $T_c$ at 3.4 K, at approximately 0.6% compression. The Young's modulus of $Sr_2RuO_4$ is 176 GPa [40], so this compression corresponds to a uniaxial pressure of ~1 GPa. The factor of 2.3 increase in $T_c$ is accompanied by more than a factor of twenty enhancement of $H_{c2}$, for fields along the *c*-axis. We complement our experimental observations with two classes of calculation. Density functional theory (DFT) calculations give evidence that the peak in $T_c$ likely coincides with a Lifshitz transition. Then, to gain insight into the effect of these large strains on possible superconducting order parameters of $Sr_2RuO_4$, we employ weak-coupling calculations that include spin-orbit and interband couplings, extending the work of Ref. [41].

Calculated band structure of $Sr_2RuO_4$

For guidance on the likely effect of strain on the electronic structure, we start with the results of the DFT calculations of the band structure of $Sr_2RuO_4$. Unstrained lattice parameters were taken from the $T = 15$ K data of Ref. [42]. In the experiment, the sample is a high-aspect-ratio bar that is compressed or tensioned along its length, so in the calculation the longitudinal strain $\varepsilon_{xx}$ is an independent variable, and the transverse strains are set, as in the experiment, according to the Poisson's ratios of $Sr_2RuO_4$: $\varepsilon_{yy} = -\nu_{xy}\varepsilon_{xx}$, and $\varepsilon_{zz} = -\nu_{xz}\varepsilon_{xx}$ [40].

The robustness of the results against different standard approximations was verified by calculations with a moderate density of $k$-points; more details are given in the Materials and Methods section. The final calculations, made in the local density approximation with spin-orbit coupling and apical oxygen position relaxation, were then extended to 343,000 $k$-points: because of proximity of the VHS to the Fermi level, an unusually large number of $k$-points was required for convergence. The first Lifshitz transition was found to occur with a compressive strain of $\varepsilon_{xx} = \varepsilon_{VHS} \approx -0.0075$. The calculated Fermi surfaces at $\varepsilon_{xx} = 0$ and $\varepsilon_{xx} = \varepsilon_{VHS}$ are shown in Fig. 1, where it can be seen that compression along $\hat{x}$ leads to a Lifshitz transition in the $\gamma$ Fermi surface along $k_y$. Thanks to the low $k_z$ dispersion, it occurs for all $k_z$ over a very narrow range of $\varepsilon_{xx}$, starting at $\varepsilon_{xx} = (-0.75 \pm 0.01) \times 10^{-2}$ and finishing by $(-0.77 \pm 0.01) \times 10^{-2}$. Cross-sections at $k_z = 0$ are also shown. In fully 2D approximations of $Sr_2RuO_4$ the Lifshitz transition occurs at a single Van Hove point, labeled in the figure and coinciding with the 2D zone boundary of an isolated $RuO_2$ sheet. The calculated change in the total density of states (DOS) as a function of tensile and compressive strains (Fig. 1C) has sharp maxima that indicate Lifshitz transitions, and should be taken as only a qualitative guide to expectations for real $Sr_2RuO_4$, in which many-body effects are likely to strengthen the quasiparticle renormalization of $v_F$ and the DOS in the vicinity of the peaks. The peak on the tension side corresponds to a Lifshitz transition along $k_x$, which is not accessible experimentally because samples break under strong tension.

Measurements of superconducting properties under uniaxial pressure

The experimental apparatus is based on that presented in Refs. [37] and [43], but modified to achieve the larger strains required for the current project. Samples were cut with a wire saw into high-aspect-ratio bars and annealed at 450 ° C for two days in air, to partially relax dislocations created by the cutting. Their ends were secured in the apparatus with epoxy [44] (Fig. 2) . Piezoelectric actuators push or pull on one end to strain the exposed central portion of the sample; to achieve high strains, 18 mm-long actuators are used, instead of the 4 mm-long ones used previously. Because samples break under strong tension, here we worked almost exclusively with compression. The superconducting transitions were measured magnetically, by measuring the mutual inductance between two coils of diameter ∼ 1 mm placed near the centre of the sample. The r.m.s. excitation field applied was ∼ 0.2 Oe, mostly parallel to the samples' $c$ axes, at frequencies between 1 and 20 kHz. Some samples also had electrical contacts for resistivity measurements.

Five samples were measured in total, and all gave consistent results. Figure 3 shows the real part of the magnetic susceptibility $\chi'$ against temperature at various compressive strains for samples 1 and 3, with zero-strain $T_c$'s of about 1.4 K. The strains are determined using a parallel-plate capacitive sensor incorporated into the apparatus. This sensor returns the applied displacement, and the sample strain is determined by dividing this displacement by the length of the strained portion of the sample. This strained length is affected in turn by elastic deformation of the epoxy that secures the sample. Comparing results from different samples, expected to have the same intrinsic behavior, yields a ~20% uncertainty in the strain determination, whose dominant origin is probably variability and uncertainty in the geometry and elastic properties of the epoxy.

When samples are initially compressed, the transition moves to higher temperature, and broadens somewhat. This broadening differs in form and magnitude from sample to sample, so is probably extrinsic. For example, imperfection in the sample mounts is likely to lead to some sample bending as force is applied, imposing a strain gradient across the thickness of the sample, and in addition a low density of dislocations and/or ruthenium inclusions may introduce some internal strain disorder. However, in spite of the likely presence of some strain inhomogeneity, the transition becomes very sharp as it approaches the maximum $T_c$, about 3.4 K. Sample 3 could be compressed well beyond this maximum, and $T_c$ was found to drop rapidly. In checks made on multiple samples, upon on releasing the strain and returning to $\varepsilon_{xx} \sim 0$, the $\chi'(T)$ curves were found to be unchanged (see Fig. S4 [45]), indicating that the sample deformation is elastic.

The peak in $T_c$ can be clearly seen in the graph of $T_c$ against $\varepsilon_{xx}$ for samples 1, 3, and 5, (Fig. 4) . The strain scales have been normalized in the plot. $\varepsilon_{xx}$ at the peak, from averaging independent determinations from samples 1, 2, 3, and 5, is $(-0.60 \pm 0.06) \times 10^{-2}$. The graph is based exclusively on magnetic measurements. The maximum $T_c$ of sample 5, at ≈3.5 K, slightly exceeds that of the other samples. Resistivity measurements can show anomalously high $T_c$ due to percolation along locally strained paths, however on samples where the resistivity was measured (samples 3 and 5), the resistive transitions never exceeded the highest magnetic $T_c$ by more than 0.08 K, confirming that it is the maximum $T_c$.

The apparatus is constructed of nonmagnetic materials, allowing measurement of the superconducting critical fields. Sample 4 was mounted in a vector magnet, with the pressure axis (a $\langle 100 \rangle$ lattice direction) parallel to the magnet $z$-axis, allowing the $c$-axis and in-plane upper critical fields to be measured in a single cool-down. The very sharp transitions in $\chi'(T)$ of $Sr_2RuO_4$ compressed to the peak in $T_c$ (referred to henceforth as $T_c = 3.4$ K $Sr_2RuO_4$) make determination of $T_c$ and $H_{c2}$ very simple: in all temperature and field ramps a sharp cusp in $\chi'(T)$ was observed, which could be identified as $T_c$ or $H_{c2}$. Specifically, the transition was identified as the intersection of linear fits to data just below and above the cusp. The in-plane $H_{c2}$ of $Sr_2RuO_4$ is known to be very sensitive to precise alignment of the field with the plane, so for in-plane measurements the vector field capability was used to align the field to within $0.2°$ of the $ab$ plane. Within the $ab$ plane, the alignment to the $\langle 100 \rangle$ direction is with standard $\sim 3°$ precision. In long field ramps the magnet was found to have $\sim 0.1$ T-scale hysteresis, so when field ramps were performed the transition was first located approximately, and then precisely with up- and down-ramps over a 0.35 T range, for which the magnet hysteresis was found to be $\sim 10$ mT.

Results are shown in Fig. 5. The $c$-axis $H_{c2}$, $H_{c2\|c}$, of $T_c = 3.4$ K $Sr_2RuO_4$ is concave, and at $T \to 0$ slightly exceeds the 1.5 T limit of the transverse coils of the vector magnet. For in-plane fields, the upper critical field $H_{c2\|a}$ reaches 4.7 T as $T \to 0$, and both temperature and field ramps show hysteresis below $\approx 1.8$ K, indicating a first-order transition.

A concave $H_{c2}(T)$ curve is an indication of high gap non-uniformity, *i.e.* substantially different gap magnitudes on different Fermi sheets, or strong variation within each sheet, or both. It has been seen in *e.g.* MgB$_2$ [46] and Be(Fe,Co)$_2$As$_2$ [47]. In $T_c = 3.4$ K Sr$_2$RuO$_4$, the slope $|dH_{c2\|c}/dT|$ is found to steadily increase to the lowest temperatures measured, although $H_{c2\|c}(T)$ must eventually become convex because $dH_{c2}/dT$ must approach zero as $T \to 0$. $H_{c2\|c}$ of unstrained Sr$_2$RuO$_4$, from Ref. [48] (Fig. 5D) is weakly concave at higher temperatures, but only above ~ 0.7 K, a much higher fraction of $T_c(H = 0)$ than the concave-convex crossover in $T_c = 3.4$ K Sr$_2$RuO$_4$. This difference in the $H_{c2}(T)$ curves indicates that the gap varies more widely across the Fermi surfaces in $T_c = 3.4$ K Sr$_2$RuO$_4$ than in unstrained Sr$_2$RuO$_4$.

Gap symmetry in $T_c = 3.4$ K Sr$_2$RuO$_4$

The $T \to 0$ critical field values for $T_c = 3.4$ K Sr$_2$RuO$_4$ are striking. $H_{c2\|c}(T \to 0)$ is enhanced by more than a factor of twenty relative to unstrained Sr$_2$RuO$_4$. $H_{c2\|a}(T \to 0)$ of unstrained Sr$_2$RuO$_4$ is 1.5 T [28], and it is enhanced by a factor of only $\approx$ 3 in $T_c = 3.4$ K Sr$_2$RuO$_4$. In the simplest picture of a fully two-dimensional triplet superconductor with the spins in the plane, the ratio $\gamma_s$ between $H_{c2\|a}$ and $H_{c2\|c}$ would be infinite, because neither orbital nor Pauli limiting would apply for in-plane fields [49]. However we observe that $\gamma_s$ is reduced from a value of $\approx$ 20 in unstrained Sr$_2$RuO$_4$ to $\approx$ 3 in $T_c = 3.4$ K Sr$_2$RuO$_4$. The electronic structure calculations presented in Fig. 1 indicate that Sr$_2$RuO$_4$ remains quasi-2D at high strains, a result supported by the observation in Fig. 5 that just below $T_c$ the slope $|dH_{c2\|a}/dT|$ far exceeds $|dH_{c2\|c}/dT|$. Therefore it seems unlikely that such a reduction in $\gamma_s$ could arise from an orbital limiting effect. In contrast, the first-order nature of the transition under strong in-plane field is consistent with a hypothesis of Pauli limiting [50], as is the absolute value of $H_{c2\|a}$. In a mean-field superconductor both $T_c$ and the Pauli-limited $H_{c2}$ are expected to vary linearly with the $T \to 0$ gap magnitude $|\Delta|$ [51]. The rise of $H_{c2\|a}$ (T->0) from 1.5 to 4.7 T in $T_c = 3.4$ K Sr$_2$RuO$_4$ is somewhat but not drastically faster than linear against $T_c$. In combination, these observations motivate investigation of whether the $T_c = 3.4$ K state might be an even-parity condensate of spin-singlet pairs.

In fact, a qualitative analysis of the enhancement of $H_{c2\|c}$ with strain also points to this possibility. In a mean-field superconductor, the orbitally-limited $H_{c2}(T \to 0)$ is proportional to a weighted average of $[|\Delta|N(E_F)]^2$, where $N(E_F)$ is the Fermi surface density of states. Because $T_c$ is proportional to a $k$-space average of $|\Delta|$, if $|\Delta(\mathbf{k})|$ is multiplied by a factor and $N(E_F)$ is not modified, the quantity $H_{c2}/T_c^2$ should remain constant. However when Sr$_2$RuO$_4$ is pressurized along a ⟨100⟩ direction $N(E_F)$ is substantially modified: it increases strongly near the Van Hove point. If $|\Delta|$ is large in this region of the Brillouin zone $H_{c2}/T_c^2$ might increase with strain. However, the Van Hove point is invariant under inversion, so $|\Delta|$ of an odd-parity order must be zero at the Van Hove point and parametrically small in its vicinity. Qualitatively, one might therefore expect stronger enhancement of $H_{c2}/T_c^2$ for even-parity order, for which large $|\Delta|$ is allowed near the Van Hove point, than for odd-parity order, where $|\Delta|$ must be small in just the regions where $N(E_F)$ is largest.

We observe, based on the data in Fig. 5, that $H_{c2\|c}(T \to 0)/T_c^2$ is enhanced by a factor of 3.6 in $T_c = 3.4$ K Sr$_2$RuO$_4$. Alternatively, because the form of $H_{c2}(T)$ is so different between unstrained and $T_c = 3.4$ K Sr$_2$RuO$_4$, it may be preferable to take a measure of $H_{c2}$ that relies only on data near $T_c$, i.e. a hypothetical $H_{c2}(0)$ for the $T \to T_c$ gap structure that excludes anomalous strengthening of the superconductivity at lower temperatures. Applying the Werthamer-Helfand-Hohenberg formula, $H_{c2}(0) = -0.7(dH_{c2}/dT)T_c$ [52], yields 0.70 and 0.056 T, respectively, for sample #4 strained to maximum $T_c$ and for the unstrained sample of Fig. 5D. If these values are used in place of the actual $H_{c2\|c}(T \to 0)$, the enhancement is 1.8. In terms of the argument discussed above, the enhancement of $\frac{H_{c2\|c}}{T_c^2}$ defined by either criterion seems to favour an even- over an odd-parity order parameter for $T_c = 3.4$ K $Sr_2RuO_4$.

To investigate these qualitative arguments in more depth and on the basis of a realistic calculation taking into account the multi-sheet Fermi surface of $Sr_2RuO_4$, we have extended to strained $Sr_2RuO_4$ a 2D weak-coupling calculation, presented in Ref. [41] as an extension of ideas first presented in Ref. [53]. The advantage of the weak-coupling approach is that it allows an unbiased comparison of different possible superconducting order parameters. Although the weak-coupling approximation is questionable in materials such as $Sr_2RuO_4$ in which the Hubbard parameter $U$ is of order the bandwidth [54], the key results of Ref. [41] were recently reproduced in a finite-$U$ calculation of $Sr_2RuO_4$ [55], further motivating the use of the weak-coupling approximation here. In our calculations, whose details are discussed further in [45], a tight-binding model of all three Fermi surfaces of $Sr_2RuO_4$ is specified, including the effects of spin-orbit and interband coupling, and fitted to the experimental dispersion. The remaining free parameter is the ratio of Hund's coupling to Hubbard interaction, $J/U$. In Ref. [41], it was found that two ranges of $J/U$ give gap anisotropy consistent with specific heat data [56]: $J/U \sim 0.08$ and $J/U \sim 0.06$. Both yield odd-parity pairing; the higher range gives helical order ($\mathbf{d} \sim p_x \hat{\mathbf{x}} + p_y \hat{\mathbf{y}}$) with $|\mathbf{d}|$ slightly larger on the $\alpha$ and $\beta$ sheets, whereas the lower value favours chiral order [$\mathbf{d} \sim (p_x \pm ip_y)\hat{\mathbf{z}}$] and $|\mathbf{d}|$ slightly larger on $\gamma$. $\mathbf{d}$ is the so-called

$d$-vector, that describes a spin-triplet order parameter, including its spin structure. For states of the type considered here, the energy gap $|\Delta|$ equals $|\mathbf{d}|$.

Here, we present $J/U = 0.06$ results for strained $Sr_2RuO_4$; the $J/U = 0.08$ results are similar [45]. At zero strain, the point group symmetry of the lattice is $D_{4h}$, and $(p_x \pm ip_y)\hat{\mathbf{z}}$ and $d_{x^2-y^2}$ are respectively the most favoured odd- and even-parity irreducible representations. At nonzero strain, the point group symmetry becomes $D_{2h}$. $(p_x \pm ip_y)\hat{\mathbf{z}}$ is resolved into the separate irreducible representations $p_x\hat{\mathbf{z}}$ and $p_y\hat{\mathbf{z}}$, and $d_{x^2-y^2}$ becomes $d_{x^2-y^2} + s$. Strain is simulated in the calculation by introducing anisotropy into the hopping integrals. The nearest-neighbor hopping $t$, for example, is resolved into $t_x = t\times(1 + a\varepsilon_{xx})$ and $t_y = t\times(1 - av_{xy}\varepsilon_{xx})$, where $a$ is chosen such that the Lifshitz transition occurs at $\varepsilon_{xx} = -0.0075$, in agreement with the LDA+SOC calculation.

$p_y\hat{\mathbf{z}}$ and $p_x\hat{\mathbf{z}}$ are respectively the highest-$T_c$ order parameters under compression and tension; compression along $\hat{\mathbf{x}}$ favors $p_y$ because it increases the density of states on the sections of Fermi surface where $p_y$ order has the largest gap magnitude, and similarly for tension and $p_x$. For $J/U = 0.06$ the possible helical orders ($\mathbf{d} \sim p_x\hat{\mathbf{x}} \pm p_y\hat{\mathbf{y}}$ or $p_x\hat{\mathbf{y}} \pm p_y\hat{\mathbf{x}}$) all have lower $T_c$ at all strains calculated. Results for $T_c$ against $\varepsilon_{xx}$ for $p_x\hat{\mathbf{z}}$, $p_y\hat{\mathbf{z}}$, and $d_{x^2-y^2} + s$ orders are shown in Fig. 6. To assign numerical values to $T_c$, the bandwidth and $U/t$ are chosen to set $T_c(\varepsilon_{xx} = 0) = 1.5$ K and $T_c(\varepsilon_{xx} = \varepsilon_{VHS}) \approx 3.4$ K; by this procedure $U/t$ comes to 6.2. $T_c$ of the $p_x$ and $p_y$ orders cross at $\varepsilon_{xx} = 0$, as they must [57], and the slope $|dT_c/d\varepsilon_{xx}|$ as $\varepsilon_{xx} \to 0$ is ~0.3 K/%. This crossing would appear as a cusp in a $T_c(\varepsilon_{xx})$ curve derived from measurements that detect only the higher $T_c$, and to search for this cusp was the primary aim of Ref. [37]. Although no cusp was seen, the resolution of that experiment does not rule out a cusp of this magnitude, and furthermore a cusp could be rounded by fluctuations [58]. At higher strains, $T_c$ of both even- and odd-parity orders is found to peak at $\varepsilon_{xx} \approx \varepsilon_{VHS}$. (The equivalent peaks on the tension side, as noted above, are not accessible experimentally.) Odd-parity order is found to be favoured at nearly all strains, however $T_c$ of the even-parity order is found to peak more strongly as the Van Hove singularity is approached, and in the immediate vicinity of the VHS even- and odd-parity orders are nearly degenerate in this calculation.

The $k$-space structure of the favored odd- and even-parity orders at $\varepsilon_{xx} = 0$ and $\varepsilon_{VHS}$ is shown in Fig. 7. For both parities, the structure of $\Delta(\mathbf{k})$ is quite complicated; $p_x \pm i p_y$, $p_y$, etc. are labels of the irreducible representation, not accurate descriptions of the full gap structure. At $\varepsilon_{xx} = \varepsilon_{VHS}$ the $p_y$ order has two nodes on the $\gamma$ sheet: one at $(0, \pi)$, where the $\gamma$ sheet touches the zone boundary and odd-parity orders must have zero amplitude, and the other along $(k_x, 0)$, where $p_y$ order has zero amplitude by symmetry. Also, whereas at zero strain the odd-parity $|\Delta|$ is generally largest on the $\gamma$ sheet, at $\varepsilon_{xx} = \varepsilon_{VHS}$ it is larger on the $\alpha$ and $\beta$ sheets, owing to the frustration for odd-parity order at the Van Hove point on the $\gamma$ sheet. $T$ still peaks at $\varepsilon_{VHS}$ because the small-$\mathbf{q}$ fluctuations on $\gamma$, which diverge at $\varepsilon_{VHS}$, also contribute to superconductivity on $\alpha$ and $\beta$ through inter-orbital interaction terms. In contrast, even-parity order does not suffer frustration at the Van Hove point. Its gap remains largest on $\gamma$ at $\varepsilon_{xx} = \varepsilon_{VHS}$, and its $T_c$ peaks more strongly.

Following Ref. [59], we calculate the orbital-limited $H_{c2\|c}/T_c^2$ at various applied strains in the semi-classical approximation. The full expression is given in [45]; an abbreviated form is: $H_{c2} \propto T_c^2 \exp(-2\langle|\psi_\mu|^2 \log \tilde{v}_\mu\rangle)$. $\langle\ldots\rangle$ is a Fermi surface average, $\psi(\mathbf{k}) \propto \Delta(\mathbf{k})$, $\mu$ is a band index, and $\tilde{v}$ is a velocity derived from the Fermi velocity. The results support the qualitative arguments made above and are shown in Fig. 8 . For $p_y$ order the shift of the gap onto the $\alpha$ and $\beta$ sheets causes a decrease in $H_{c2\|c}/T_c^2$, because these sheets have lower DOS than the $\gamma$ sheet. In contrast, the increased DOS around the Van Hove point causes $H_{c2\|c}/T_c^2$ of $d_{x^2-y^2} + s$ order to increase towards the VHS. The actual $H_{c2\|c}/T_c^2$ may be enhanced over the weak-coupling results by strengthened many-body effects towards the VHS, however the results emphasize a strong quantitative disparity between $H_{c2\|c}/T_c^2$ for even- and odd-parity order parameters.

We note that if unstrained $Sr_2RuO_4$ has $p_x \pm ip_y$ order, at nonzero strain the low-$T$ order is likely still to be chiral, but with different amplitudes of the $p_x$ and $p_y$ components. In Fig. 8, the goal is to determine the expected trend in $H_{c2\|c}/T_c^2$ for odd-parity order by comparing the same irreducible representation, $p_x$ or $p$ , at different strains. If the order is actually $ap_x \pm ibp_y$, with $a \neq b$, $H_{c2\|c}$ will generally be higher, but a similar trend in $H_{c2\|c}/T_c^2$ is expected.

Although heat capacity data suggest $J/U \sim 0.06$ or $\sim 0.08$, we also considered $J/U$ over a wider range, from 0 to 0.3. The essential qualitative features presented here for $J/U = 0.06$, the peak in $T_c$ at the Lifshitz transition for both even- and odd-parity order, and the enhancement (suppression) of $H_{c2}/T_c^2$ for even (odd) parity, are found to occur across this range. Results for $J/U = 0.08, 0$, and $0.25$ are shown in [45].

**Discussion**

One long-standing puzzle in the physics of $Sr_2RuO_4$ has been the origin of the so-called 3 K phase, which is $T_c \sim 3$ K superconductivity observed in eutectic crystals containing inclusions of Ru metal in a matrix of $Sr_2RuO_4$ [60]. It has been established that this higher-$T_c$ superconductivity has a low volume fraction [60, 61], showing that it occurs at the inclusions rather than the bulk, and further that it occurs on the $Sr_2RuO_4$ side of Ru-$Sr_2RuO_4$ interfaces [62]. Although full proof would require observation of the strain field around Ru inclusions, it now seems very likely that local internal strain is the origin of the 3 K phase. The upper critical fields of the 3 K phase have been obtained through measurement of resistivity along extended inclusions, and were found to be ~1 T for $c$-axis and ~3.5 T for in-plane fields [63]. The similarity of these fields with the critical fields of bulk $T_c = 3.4$ K $Sr_2RuO_4$ further supports the hypothesis that the 3 K phase is a local strain effect, although it is also possible that the observed 3 K phase critical fields are enhanced by the two-dimensional geometry of interface superconductivity [63, 64].

Three-band models in Refs. [55] and [65], in addition to the calculations presented here, identify the proximity of the $\gamma$ sheet to a VHS as an important factor in the superconductivity of $Sr_2RuO_4$. Simultaneous to this work, calculations in Refs. [66] and [67] have found increasing $T_c$, at least initially, on tuning towards the VHS with strain. That the peak in $T_c$ occurs at a similar strain to $\varepsilon_{VHS}$ determined from DFT calculations suggests that it coincides with the Lifshitz transition. However an alternative possibility is that $T_c$ of an odd-parity order initially increases, thanks to the increase in DOS induced by compression, but then decreases as frustration at the Van Hove point becomes more important. This is not the behavior indicated by our calculations, where $T_c$ of $p_y$ order peaks at $\varepsilon_{VHS}$, but may still be considered a qualitative possibility. A further possibility, from Ref. [67], is that compression stabilizes competing spin density wave order that cuts off the superconductivity before $\varepsilon_{VHS}$.

Evidence that the $T_c$ peak and Lifshitz transition do in fact coincide comes from preliminary transport data. In the normal state, inelastic scattering is generally expected to scale with the Fermi level density of states, so at nonzero temperature a peak in the resistivity at the Lifshitz transition is expected. The resistivity $\rho_{xx}$ at 4.5 K, above the highest $T_c$, indeed peaks in the vicinity of the $T_c$ peak (Fig. S3). At higher strains it falls rapidly, to below its zero-strain value. The calculated Fermi surface density of states (Fig. 1 C) similarly drops to below its zero-strain value beyond $\varepsilon_{VHS}$. The resistivity does not show the sharp increase generically expected with transitions into phases involving a gap. Further experiments are needed to determine the precise behavior of the normal-state resistivity across the $T_c$ peak.

Although important, the issue of whether the peak in $T_c$ coincides with the Lifshitz point does not strongly affect the main conclusions that we draw here, because the substance of the comparison of the critical fields of $T_c = 3.4$ K and unstrained $Sr_2RuO_4$ stands regardless. The weak-coupling calculations yield strongly divergent trends for $H_{c2\|c}/T_c^2$ for even- and odd-parity order at all intermediate strains, not only at the VHS, and because this is a result of frustration of odd-parity order in the vicinity of the Van Hove point it is unlikely to be strongly model-dependent. Also, the arguments for Pauli limiting of $H_{c2\|a}(T \to 0)$ are unaffected by whether the peak is at the Lifshitz transition. The critical field comparisons clearly raise the possibility that the $T_c = 3.4$ K superconductivity has an even-parity, spin-singlet order parameter. It is difficult to understand in a naive analysis how the critical field anisotropy $\gamma_s$ could be only $\approx 3$ without Pauli limiting of $H_{c2\|a}$. However most current theories of $Sr_2RuO_4$ are two-dimensional and make no predictions for $\gamma_s$; we believe our observations provide strong motivation for extending realistic three-band calculations into the third dimension.

If the 3.4 K superconducting state is even-parity, there are two obvious possibilities, both exciting, for its relationship with the superconductivity of unstrained $Sr_2RuO_4$. One is that the evolution of the order parameter is continuous between the two states, and unstrained $Sr_2RuO_4$ is also an even-parity superconductor. The appearance of a first-order transition at low temperatures for in-plane fields in both $T_c = 3.4$ K (Fig. 5 A) and unstrained $Sr_2RuO_4$ [27] also argues for this possibility. However in this case a substantial body of experimental evidence [30] for triplet, chiral order would require alternative explanation. The evidence for chirality could be accommodated by a spin-singlet state, $d_{xz} \pm id_{yz}$ [68]. This order parameter has horizontal line nodes, which requires interplane pairing, and would be surprising in such a highly two-dimensional material as $Sr_2RuO_4$. However it would, again, be useful to extend calculations into the third dimension so that it could be compared on an equal footing with the more standard candidate order parameters based on intraplane pairing. The other possibility is that there is a transition at an intermediate strain between odd- and even-parity states. At such a transition a kink, possibly weak, is expected in $T_c(\varepsilon_{xx})$, and a jump in $H_{c2\|c}(T \rightarrow 0)$. An important follow-up experiment therefore is measurement of $H_{c2\|c}$ at intermediate strains. This has not been done yet because the broadening of the transitions at intermediate strains complicates accurate determination of $H_{c2}$, and higher-precision sample mounting methods may be required.

Consideration of an odd-to-even-parity transition at intermediate strains is also motivated by evidence for interference between the superconductivity of Ru inclusions and that of bulk $Sr_2RuO_4$, and for hysteresis and switching behavior in Ru/$Sr_2RuO_4$ systems. The possible interference appears as a sharp drop in the critical current $I_c$ of Pb/Ru/$Sr_2RuO_4$ junctions at

$T_c$ of $Sr_2RuO_4$ [69, 70], which has been interpreted as an onset of phase frustration at the Ru/SRO interface. However it could perhaps also be explained by appearance of an odd-parity/even-parity interface around the Ru inclusion. Similarly, hysteretic $I_c$ has been seen in $Sr_2RuO_4$/Cu/Pb [18], Nb/Ru/$Sr_2RuO_4$ [71], and Pb/Ru/$Sr_2RuO_4$ [70] junctions, and microbridges of $Sr_2RuO_4$ with Ru inclusions [72]. The former two also showed time-dependent switching noise. All these results have been interpreted as motion of $p_x + ip_y/p_x - ip_y$ domain walls, however even/odd domain walls appear to be a viable alternative possibility.

Our observations also give cause for optimism concerning the prospects of finding superconductivity in biaxially strained thin films: the factor-of-twenty $H_{c2\|c}$ enhancement corresponds to a factor of 4.5 reduction in the coherence length, considerably reducing the disorder constraint for unconventional superconductivity. Biaxial lattice expansion preserves tetragonal symmetry and induces Lifshitz transitions at the $X$ and $Y$ Van Hove points simultaneously, and so may induce qualitatively different superconductivity than tuning to a single Van Hove point with uniaxial pressure.

Finally, our results provide strong motivation for extending the application of piezoelectric-based strain tuning to other materials. In this work we have demonstrated that compressions up to ~1% are possible, with in situ tunability and good strain homogeneity. The fact that we have achieved a factor of 2.3 increase of $T_c$ of an unconventional superconductor points the way to substantial tuning of properties of other material classes as well.

### Materials and Methods

Relativistic DFT electronic structure calculations were performed using the full-potential local orbital FPLO code [73, 74, 75], version fplo14.00-49. For the exchange-correlation potential, within the local density (LDA) and the the general gradient approximation (GGA) the parametrizations of Perdew-Wang [76] and Perdew-Burke-Ernzerhof [77] were chosen, respectively. The spin-orbit coupling (SOC) was treated non-perturbatively solving the four component Kohn-Sham-Dirac equation [78]. Initial calculations were performed on 8000 $k$-points (20×20×20 mesh), both in the LDA and GGA approximations, with and without SOC, and with and without apical oxygen relaxation. All these calculations gave similar results, with the calculated $\varepsilon_{VHS}$ between -0.012 (GGA + relaxation) and -0.009 (LDA+SOC+relaxation). However proximity of the VHS to the Fermi level meant that convergence to within 3% of the calculated energy of the VHS to $E_F$ required a higher density of $k$-points, so LDA+SOC+relaxation calculations were then carried out on a mesh of 343,000 $k$-points (70x70x70 mesh, 44766 points in the irreducible wedge of the Brillouin zone), placing $\varepsilon_{VHS}$ at -0.0075.

Although we believe that using experimentally determined structural parameters for unstrained Sr$_2$RuO$_4$ (as described in the main text) is the most natural starting point for the calculations, we also checked for the effect of fully relaxing the structure in the local density approximation. That relaxation only slightly reduced the cell volume (by 2.7%), preserved the $c/a$ ratio to within 0.1% and led to an increase of only 0.001 in $\varepsilon_{VHS}$, so we are confident that use of a relaxed structure gives no substantial systematic change compared to use of the experimental one.

The pressure apparatus is based on that described in Ref. [43], however there are a few key modifications that merit mention here. (1) The piezoelectric actuators were 18 mm-long Physik Instrumente PICMA linear actuators. (2) The displacement sensor is a parallel-plate capacitor, in place of the strain gauge described in Refs. [37] and [43]. The data in this work suggest that the strains determined in Ref. [37] are ≈35% too low. One very likely contribution to this error is the mechanical resistance imposed by the strain gauge on the motion of the original apparatus. Temperature shifts in the gauge coefficient of the strain gauge may also contribute. Capacitive sensors are less affected by field and temperature, and impose no mechanical resistance, so we have more confidence in the strains reported in this work. (3) The thermal contraction foils have been eliminated, allowing the core of the apparatus to be made as a single piece. The longer actuators have more than sufficient range to overcome differential thermal contraction between the sample and apparatus.

When mounting samples, a small voltage is often applied to the actuators to move the sample mount points slightly further apart. When this voltage is later released the sample is placed under modest compression. This step reduces the risk that the sample will break during cooling, for example if temperature inhomogeneity in the apparatus places the sample under inadvertent tension.

To estimate the strain applied to a sample, two pieces of information are required. The first is the origin of the strain scale, the point where the sample is under zero strain. In Ref. [37] it was determined that $T_c$ of Sr$_2$RuO$_4$ is minimum within experimental error at zero strain, so for most samples the origin can be identified as the minimum in $T_c$. Samples 1 and 4 broke during cooling, and could be compressed by closing the crack, but not tensioned. The process of re-engaging the two ends can be gradual, *e.g.* if the two faces of the crack do not match perfectly, so zero strain cannot be reliably identified by attempting to locate a precise point where $T_c(\varepsilon_{xx})$ starts to change. Instead, a quadratic fit was made to the $T_c(\varepsilon_{xx})$ curve over a temperature range near but above the lowest observed $T_c$. Zero strain was identified as the minimum of the fitted curve, plus $2 \cdot 10^{-4}$ to account for the anomalous flattening of $T_c(\varepsilon_{xx})$ around $\varepsilon_{xx} = 0$ observed in Ref. [37]. The other piece of information required is an effective strained length: the capacitive sensor measures a displacement, and $\varepsilon_{xx}$ is this displacement divided by the effective strained length. Deformation of the sample mounting epoxy means that the effective strained length is typically ~0.4 mm longer than the exposed length of the sample. It is estimated through finite element analysis, as described in Refs. [37] and [43].

The layers of the epoxy that secure the sample are generally 20–40 $\mu$m thick, an estimated broad optimum. Thinner layers transmit force to the sample more efficiently (*i.e.* give a shorter effective strained length), while thicker layers reduce stress concentration in the epoxy and allow greater tolerance in assembly. The dimensions, calculated effective strained length, and estimated $\varepsilon_{xx}$ at the peak in $T_c$ for each sample are given in [45].

**Acknowledgements**

We thank Heike Pfau for experimental contributions, Eun-Ah Kim, Steven Kivelson, Srinivas Raghu, Kyle Shen and Fu-Chun Zhang for stimulating discussions, and Eun-Ah Kim and Fu-Chun Zhang for sharing the results of their calculations with us. Kim and collaborators (Ref. [66]) used renormalization group calculations to study $T_c$ versus uniaxial but mainly biaxial strain, while Liu, Zhan, Rice and Wang (Ref. [67]) employed functional renormalization group calculations of the strain dependence of $T_c$ concentrating on the $d_{xy}$-based Fermi surface sheet. On topics where they overlap, the reults of those two calculations, as well as the the calculations presented in this paper, are qualitatively similar. We gratefully acknowledge the support of the Max Planck Society and the UK EPSRC under grants EP/1031014/1, EP/G03673X/1, EP/N01930X/1, and EP/I032487/1. LZ acknowledges the support of the China Scholarship Council. TS acknowledges the support of the Clarendon Fund Scholarship, the Merton College Domus and Prize Scholarships, and the University of Oxford. YM acknowledges support by the JSPS Grant-in-Aids on Topological Quantum Phenomena (KAKENHI JP22103002) and on Topological Materials Science (KAKENHI JP15H05852). CH has 31% ownership of Razorbill Instruments, which has commercialized apparatus based on that used in this work. Raw data for all figures in this paper are available at ***.

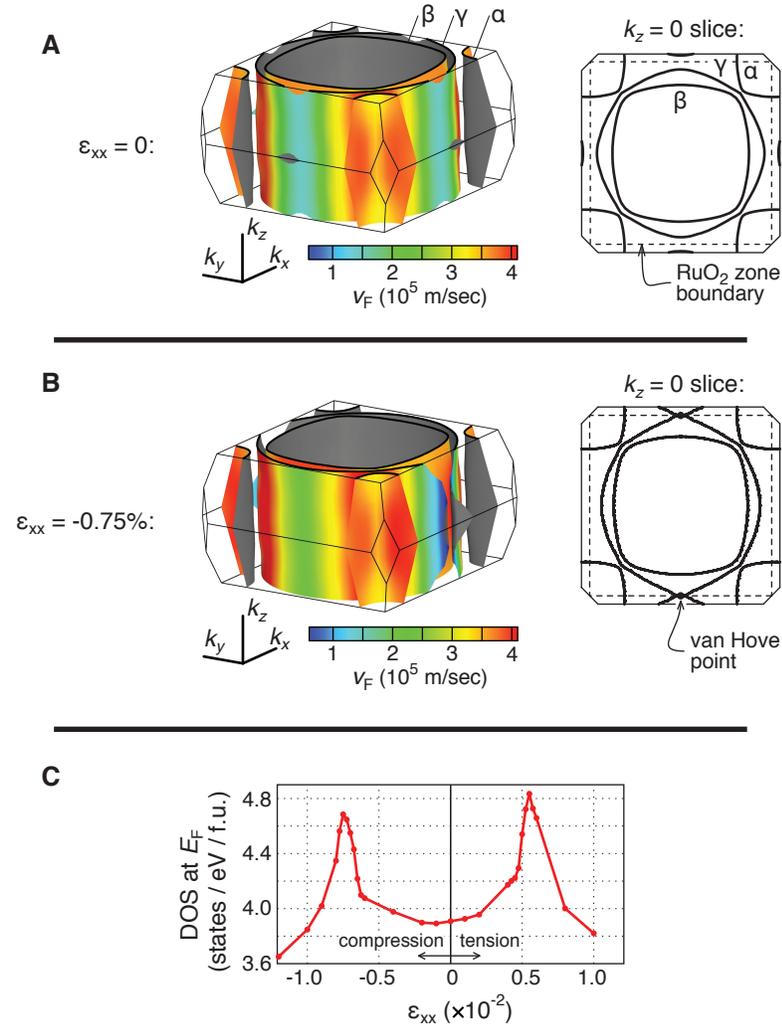

Figure 1: **DFT calculation results.** (**A**) Calculated Fermi surfaces of unstrained $Sr_2RuO_4$, colored by the Fermi velocity $v_F$, at zero strain. The three surfaces are labeled $\alpha$, $\beta$, and $\gamma$. A cross-section through $k_z = 0$ is also shown. The dashed lines indicate the zone of an isolated $RuO_2$ sheet; in 2D models of $Sr_2RuO_4$, the Van Hove point is located on this zone boundary. **(B)** Calculated Fermi surfaces at $\varepsilon_{xx} = -0.0075$. **(C)** Calculated total density of states against $\varepsilon_{xx}$.

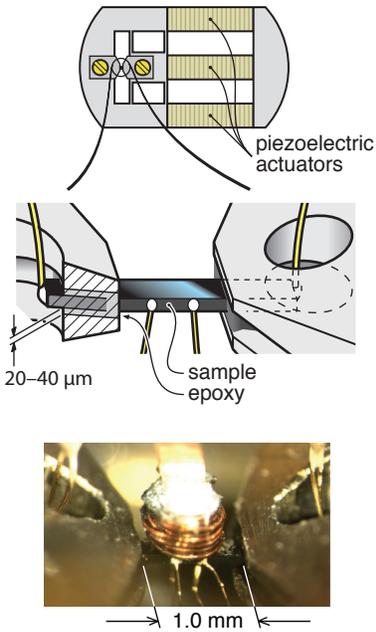

Figure 2: **Apparatus and sample configuration.** Top: apparatus configuration. Extending the outer two piezoelectric actuators tensions the sample, and extending the central actuator compresses the sample. Middle: sample configuration. The ends are secured with epoxy. Some samples have contacts (shown schematically) for resistivity measurements. Bottom: a photograph of sample 3. On top of the sample, mounted on a flexible cantilever, are concentric coils used for measuring magnetic susceptibility.

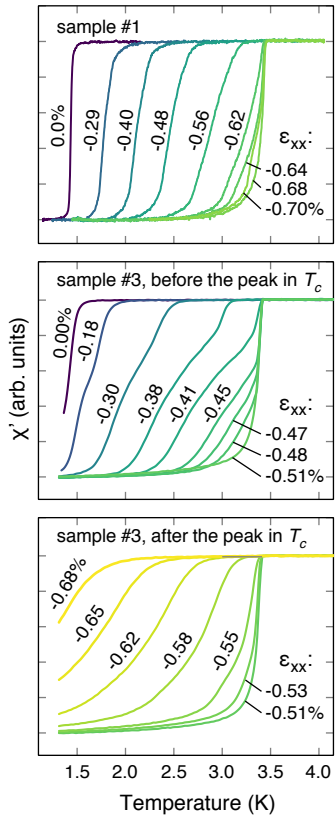

Figure 3: **Susceptibility against temperature.** Top: real part of the susceptibility $\chi$ against temperature for sample 1, at various $\varepsilon_{xx}$. No normalizations or offsets are applied to the curves. Middle and bottom: same, for sample 3.

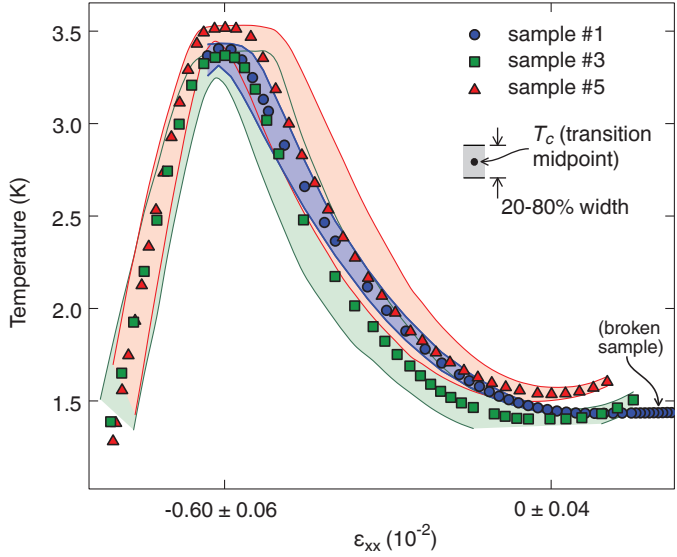

Figure 4: **$T_c$ against strain for samples 1, 3, and 5.** The points are the midpoints (50% levels) of the transitions shown in Fig. 3 , and the lines are the 20 and 80% levels, giving a measure of the transition width. The strain scales have been normalized. We estimate an uncertainty of $0.04 \times 10^{-2}$ on the determination of zero strain of each sample, and the strain at the peak in $T_c$ is determined by averaging independent determinations from four samples to be $(-0.60 \pm 0.06) \times 10^{-2}$.

The flat region around $\varepsilon_{xx} = 0$ for sample 1 is an artefact: the sample broke during cool-down, meaning that tensile strain could not be applied, and a compressive displacement was required for it to re-engage.

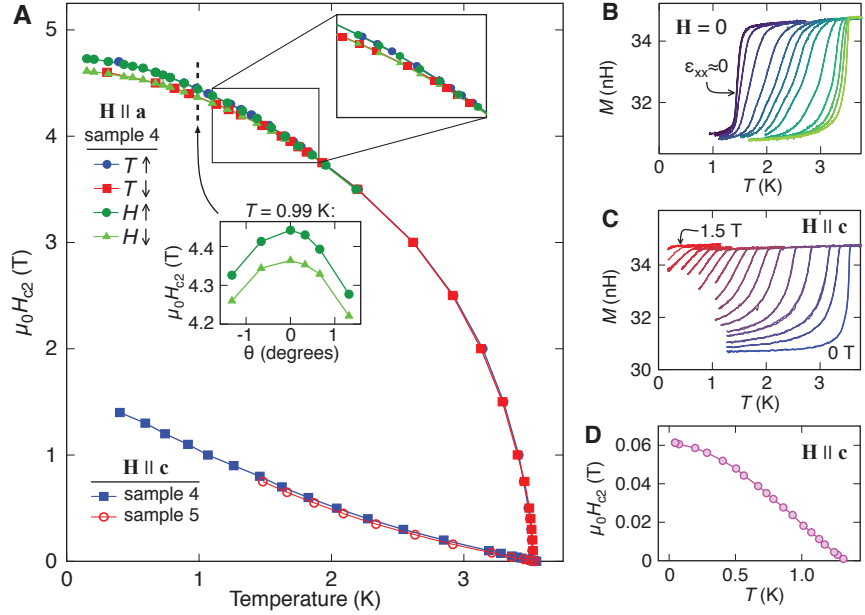

Figure 5: **$H_{c2}$ against temperature.** (A) $H_{c2\|a}$ and $H_{c2\|c}$ against temperature for sample 4, compressed to the peak in $T_c$. $H_{c2\|a}$ was measured with both field and temperature ramps, and found to be hysteretic below ~1.8 K (upper inset). Lower inset: angle dependence of $H_{c2}$ at 990 mK, confirming the field alignment. $\theta$ is the angle in the **a-c** plane, and $\theta = 0$ the field angle at which the $H_{c2\|a}$ data were collected. (B) Raw data for $\chi'(T)$ of sample 4 at various $\varepsilon_{xx}$. The $y$ axis is the mutual inductance of the measurement coils. (C) Measured $\chi'(T)$ at the peak in $T_c$, and at fields in 0.1 T increments between 0 and 1.5 T. (D) Data for $H_{c2\|c}$ of an unstrained $Sr_2RuO_4$ sample with slightly sub-optimal $T_c$, from Ref. [48].

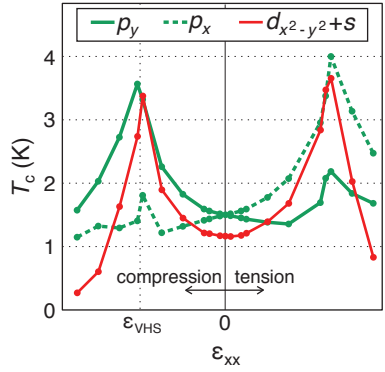

Figure 6: **Weak-coupling calculations: $T_c$ versus strain.** The bandwidth and $U/t$ were set to reproduce the experimental values of $T_c(0) = 1.5$ K, and $T_c(\varepsilon_{\mathrm{VHS}}) \sim 3.4$ K. $p_x \pm ip_y$ and $d_{x^2-y^2}$ are irreducible representations of the $\varepsilon_{xx} = 0$ (i.e. tetragonal) lattice. For $\varepsilon_{xx} \neq 0$, $p_x \pm ip_y$ is resolved into separate representations $p_x$ and $p_y$, and $d_{x^2-y^2}$ becomes $d_{x^2-y^2} + s$.

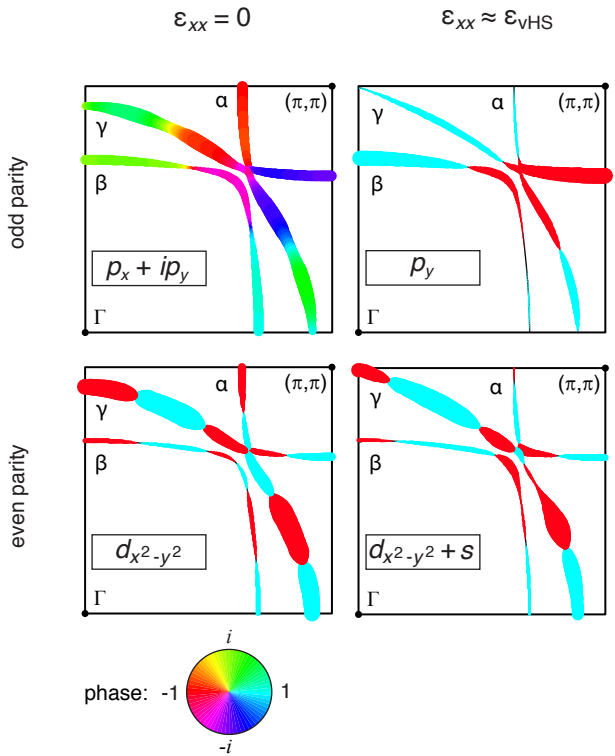

Figure 7: **Weak-coupling calculations: order parameters.** Top: The odd-parity order parameter at $\varepsilon = 0$ and $\varepsilon \approx \varepsilon_{\text{VHS}}$; the VHS is reached at $(0, \pi)$. The width of the traces is proportional to the energy gap, and the color indicates the phase. For $\varepsilon_{xx} \neq 0$, $p_x \pm ip_y$ is no longer an irreducible representation of the lattice, so the $p_y$ representation, the favoured order parameter for $\varepsilon_{xx} < 0$, is shown instead. Bottom: even-parity order, at $\varepsilon = 0$ and $\varepsilon \approx \varepsilon_{\text{VHS}}$.

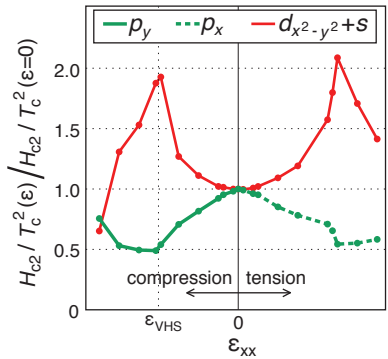

Figure 8: **Weak-coupling calculations: $H_{c2\|c}/T_c^2$ versus strain.** The results are normalized to the $H_{c2\|c}/T_c^2$ calculated at zero strain.

Supplementary Materials

The Supplementary Materials include:
Supplementary Text
    Additional Data
    Further details on the weak-coupling calculation
    Results of the weak-coupling calculation at other values of *J/U*
    Considerations in three dimensions
Figures S1 to S7
Table S1
References 81 to 87